\documentclass[allclo, onecollarge, square, comma, sort&compress, natbib]{svjour2}
\bibpunct{[}{]}{;}{n}{}{,} 
\smartqed  
\usepackage{amsmath}
\usepackage{amssymb}
\usepackage{graphicx}
\usepackage{color}

\definecolor{darkgreen}{rgb}{0,0.5,0}
\definecolor{purple}{rgb}{0.5,0,0.5}
\definecolor{blue}{rgb}{0.0,0.0,0.50}
\definecolor{scarlet}{rgb}{1.0,0.2,0}

\usepackage[mathscr,scaled=1.15]{urwchancal}
\DeclareFontFamily{OT1}{pzc}{}
\DeclareFontShape{OT1}{pzc}{m}{it}%
{<-> s * [1.15] pzcmi7t}{}
\DeclareMathAlphabet{\mathpzc}{OT1}{pzc}{m}{it}

\journalname{Few-Body Systems}
\begin{document}

\title{Baryons and the Borromeo}


\author{Craig D.\ Roberts \and Jorge Segovia}


\institute{Craig D.\ Roberts \at
            Physics Division, Argonne National Laboratory, Argonne, Illinois 60439, USA.
              \email{cdroberts@anl.gov}
\and
    Jorge Segovia \at
Physik-Department, TU-M\"unchen,
D-85748 Garching, Germany
\email{jorge.segovia@tum.de}}

\date{Received: 11 February 2016 / Accepted: 08 March 2016 }

\maketitle

\begin{abstract}
The kernels in the tangible matter of our everyday experience are composed of light quarks.  At least, they are light classically; but they don't remain light.  Dynamical effects within the Standard Model of Particle Physics change them in remarkable ways, so that in some configurations they appear nearly massless, but in others possess masses on the scale of light nuclei.  Modern experiment and theory are exposing the mechanisms responsible for these remarkable transformations.  The rewards are great if we can combine the emerging sketches into an accurate picture of confinement, which is such a singular feature of the Standard Model; and looming larger amongst the emerging ideas is a perspective that leads to a Borromean picture of the proton and its excited states.
\end{abstract}

\section{Introduction}
\label{secConfinement}
Quantum chromodynamics (QCD) is a local, relativistic, non-Abelian, quantum gauge-field theory, which possesses the property of asymptotic freedom, \emph{i.e}.\ QCD interactions are weaker than Coulombic at short distances.  This behaviour is evident in the one-loop expression for the running coupling, $\alpha_s(Q^2)$, and verified in a host of experiments.  Hence, as a necessary consequence of asymptotic freedom, $\alpha_s(Q^2)$ must increase as $Q^2/\Lambda_{\rm QCD}^2\to 1^+$, where $\Lambda_{\rm QCD} \sim 200\,$MeV is the natural mass-scale of QCD, whose dynamical generation through quantisation spoils the conformal invariance of the classical massless theory \cite{Collins:1976yq, Nielsen:1977sy, tarrach}.  In fact, at $Q^2\approx 4\,$GeV$^2=: \zeta_2^2$, which corresponds to a length-scale on the order of 10\% of the proton's radius, it is empirically known that $\alpha_s(\zeta_2^2)\gtrsim 0.3$.  These observations describe a peculiar circumstance, \emph{viz}.\ an interaction that becomes stronger as the participants try to separate.  It leads one to explore some curious possibilities: If the coupling grows so strongly with separation, then perhaps it is unbounded; and perhaps it would require an infinite amount of energy in order to extract a quark or gluon from the interior of a hadron?  Such thinking has led to the \\[1ex]
\hspace*{3em}\parbox[t]{0.9\textwidth}{\textit{Confinement Hypothesis}: Colour-charged particles cannot be isolated and therefore cannot be directly observed.  They clump together in colour-neutral bound-states.}\\

Confinement seems to be an empirical fact; but a mathematical proof is lacking.  Partly as a consequence, the Clay Mathematics Institute offered a ``Millennium Problem'' prize of \$1-million for a proof that $SU_c(3)$ gauge theory is mathematically well-defined \cite{Jaffe:Clay}, one consequence of which will be an answer to the question of whether or not the confinement conjecture is correct in pure-gauge QCD.

There is a problem with that, however: no reader of this article can be described within pure-gauge QCD.  The presence of quarks is essential to understanding all known visible matter, so a proof of confinement which deals only with pure-gauge QCD is chiefly irrelevant to our Universe.  We exist because Nature has supplied two light quarks; and those quarks combine to form the pion, which is unnaturally light ($m_\pi<\Lambda_{\rm QCD}$) and hence very easily produced.

One may bring this arcanum into sharper focus by noting that one aspect of the Yang-Mills millennium problem \cite{Jaffe:Clay} is to prove that pure-gauge QCD possesses a mass-gap $\Delta>0$.  There is strong evidence supporting this conjecture, found especially in the fact that numerical simulations of lattice-regularised QCD (lQCD) predict $\Delta \gtrsim 1.5\,$GeV \cite{McNeile:2008sr}.  However, with $\Delta^2/m_\pi^2 \gtrsim 100$, can the mass-gap in pure Yang-Mills really play any role in understanding confinement when dynamical chiral symmetry breaking (DCSB), very likely driven by the same dynamics, ensures the existence of an almost-massless strongly-interacting excitation in our Universe?  If the answer is not \emph{no}, then it should at least be that one cannot claim to provide a pertinent understanding of confinement without simultaneously explaining its connection with DCSB.  The pion must play a critical role in any explanation of confinement in the Standard Model; and any discussion that omits reference to the pion's role is \emph{practically} \emph{irrelevant}.

This perspective is canvassed elsewhere \cite{Cloet:2013jya} and can be used to argue that the potential between infinitely-heavy quarks measured in numerical simulations of quenched lQCD -- the so-called static potential \cite{Wilson:1974sk} -- is disconnected from the question of confinement in our Universe.  This is because light-particle creation and annihilation effects are essentially nonperturbative in QCD, so it is impossible in principle to compute a quantum mechanical potential between two light quarks \cite{Bali:2005fuS, Prkacin:2005dc, Chang:2009ae}.  Consequently, there is no measurable flux tube in a Universe with light quarks and hence the classical flux tube cannot be the correct paradigm for confinement.

As highlighted already, DCSB is the key here.  It ensures the existence of pseudo-Nambu-Goldstone modes; and in the presence of these modes, no flux tube between a static colour source and sink can have a measurable existence.  To verify this, consider such a tube being stretched between a source and sink.  The potential energy accumulated within the tube may increase only until it reaches that required to produce a particle-antiparticle pair of the theory's pseudo-Nambu-Goldstone modes.  Simulations of lQCD show \cite{Bali:2005fuS, Prkacin:2005dc} that the flux tube then disappears instantaneously along its entire length, leaving two isolated colour-singlet systems.  The length-scale associated with this effect in QCD is $r_{\not\sigma} \simeq (1/3)\,$fm and hence if any such string forms, it would dissolve well within a hadron's interior.

An alternative realisation associates confinement with dramatic, dynamically-driven changes in the analytic structure of QCD's propagators and vertices.  That leads coloured $n$-point functions to violate the axiom of reflection positivity and hence forces elimination of the associated excitations from the Hilbert space associated with asymptotic states \cite{GJ81}.  This is certainly a sufficient condition for confinement \cite{Stingl:1985hx, Krein:1990sf, Hawes:1993ef, Roberts:1994dr}.  It should be noted, however, that the appearance of such alterations when analysing some truncation of a given theory does not mean that the theory itself is truly confining: unusual spectral properties can be introduced by approximations, leading to a truncated version of a theory which is confining even though the complete theory is not, \emph{e.g}.\ Refs.\,\cite{Krein:1993jb, Bracco:1993cy}.  Notwithstanding such exceptions, a computed violation of reflection positivity by coloured functions in a veracious treatment of QCD does express confinement.  Moreover, via this mechanism, it is achieved as the result of an essentially dynamical process.  It is known that both quarks and gluons acquire a running mass in QCD \cite{Bhagwat:2003vw, Bowman:2005vx, Bhagwat:2006tu, Aguilar:2008xm, Aguilar:2009nf, Boucaud:2011ugS, Pennington:2011xs, Ayala:2012pb, Binosi:2014aea}; and the generation of these masses leads to the emergence of a length-scale $\varsigma \approx 0.5\,$fm, whose existence and magnitude is evident in all existing studies of dressed-gluon and -quark propagators, and which characterises the striking change in their analytic structure that has just been described.  In models based on such features \cite{Stingl:1994nk}, once a gluon or quark is produced, it begins to propagate in spacetime; but after each ``step'' of length $\varsigma$, on average, an interaction occurs so that the parton loses its identity, sharing it with others.  Finally a cloud of partons is produced, which coalesces into colour-singlet final states.  This picture of parton propagation, hadronisation and confinement can be tested in experiments at modern and planned facilities \cite{Accardi:2009qv, Dudek:2012vr, Accardi:2012qutS}.

\section{Enigma of mass}
\begin{figure}[t]
\leftline{\includegraphics[clip,width=0.45\textwidth]{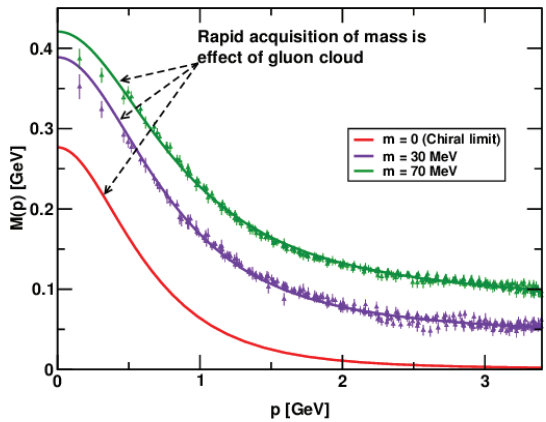}}
\vspace*{-40ex}

\rightline{\parbox{24em}{
\caption{\label{gluoncloud}
Mass function, $M(p)$, a characterising feature of the dressed-quark propagator in Eq.\,\eqref{Spgen}.  \emph{Solid curves} -- DSE results, explained in Refs.\,\protect\cite{Bhagwat:2003vw,Bhagwat:2006tu}, ``data'' -- numerical simulations of lattice-regularised QCD (lQCD) \protect\cite{Bowman:2005vx}.  (NB.\ $m=70\,$MeV is the uppermost curve and current-quark mass decreases from top to bottom.)  The current-quark of perturbative QCD evolves into a constituent-quark as its momentum becomes smaller.  The constituent-quark mass arises from a cloud of low-momentum gluons attaching themselves to the current-quark.  This is DCSB: an essentially nonperturbative effect that generates a quark \emph{mass} \emph{from nothing}; namely, it occurs even in the chiral limit.}}}
\vspace*{6ex}

\end{figure}

DCSB is a crucial emergent phenomenon in QCD.  It is expressed in hadron wave functions, not in vacuum condensates \cite{Brodsky:2009zd, Brodsky:2010xf, Chang:2011mu, Brodsky:2012ku, Cloet:2013jya}; and contemporary theory argues that DCSB is responsible for more than 98\% of the visible mass in the Universe.  Given that classical massless-QCD is a conformally invariant theory, this means that DCSB is the origin of \emph{mass from nothing}.  This effect is evident in the dressed-quark propagator:
\begin{equation}
\label{Spgen}
S(p) = 1/[i \gamma\cdot p A(p^2) + B(p^2)] = Z(p^2)/[i\gamma\cdot p + M(p^2)]\,,
\end{equation}
where $M(p^2)$ is the dressed-quark mass-function, the behaviour of which is depicted and explained in Fig.\,\ref{gluoncloud}.  It is important to insist on the term ``dynamical,'' as distinct from spontaneous, because nothing is added to QCD in order to effect this remarkable outcome and there is no simple change of variables in the QCD action that will make it apparent.  Instead, through the act of quantising the classical chromodynamics of massless gluons and quarks, a large mass-scale is generated.

DCSB is very clearly revealed in properties of the pion, whose structure is described by a Bethe-Salpeter amplitude:
\begin{equation}
\Gamma_{\pi}(k;P) = \gamma_5 \left[
i E_{\pi}(k;P) + \gamma\cdot P F_{\pi}(k;P)  + \gamma\cdot k \, G_{\pi}(k;P) + \sigma_{\mu\nu} k_\mu P_\nu H_{\pi}(k;P)
\right],
\label{genGpi}
\end{equation}
where $k$ is the relative momentum between the  valence-quark and -antiquark constituents (defined here such that the scalar functions in Eq.\,\eqref{genGpi} are even under $k\cdot P \to - k\cdot P$) and $P$ is their total momentum.  $\Gamma_{\pi}(k;P)$ is simply related to an object that would be the pion's Schr\"odinger wave function if a nonrelativistic limit were appropriate.  In QCD if, and only if, chiral symmetry is dynamically broken, then one has in the chiral limit \cite{Maris:1997hd, Qin:2014vya}:
\begin{equation}
\label{gtrE}
f_\pi E_\pi(k;0) = B(k^2)\,.
\end{equation}
This identity is miraculous.\footnote{Eq.\,\eqref{gtrE} has many corollaries, \emph{e.g}.\ it ensures that chiral-QCD generates a massless pion in the absence of a Higgs mechanism; predicts $m_\pi^2 \propto m$ on $m \simeq 0$, where $m$ is the current-quark mass; and entails that the chiral-limit leptonic decay constant vanishes for all excited-state $0^-$ mesons with nonzero isospin \cite{Holl:2004fr, Ballon-Bayona:2014oma}.}
It is true in any covariant gauge, independent of the renormalisation scheme; and it means that the two-body problem is solved, nearly completely, without lifting a finger, once the solution to the one body problem is known.  Eq.\,\eqref{gtrE} is a quark-level Goldberger-Treiman relation.  It is also the most basic expression of Goldstone's theorem in QCD, \emph{viz}.\\[-3ex]
\hspace*{3em}\parbox[t]{0.9\textwidth}{\flushleft \emph{Goldstone's theorem is fundamentally an expression of equivalence between the one-body problem and the two-body problem in QCD's colour-singlet pseudoscalar channel}.}
\smallskip

\hspace*{-\parindent}Consequently, pion properties are an almost direct measure of the dressed-quark mass function depicted in Fig.\,\ref{gluoncloud}; and the reason a pion is massless in the chiral limit is simultaneously the explanation for a proton mass of around 1\,GeV.  Thus, enigmatically, properties of the nearly-massless pion are the cleanest expression of the mechanism that is responsible for almost all the visible mass in the Universe.

\section{Continuum-QCD and \emph{ab initio} predictions of hadron observables}
\label{secAbInitio}
\begin{figure}[t]
\rightline{\begin{minipage}[t]{1.0\textwidth}
\begin{minipage}{0.47\textwidth}
\centerline{\includegraphics[clip,width=0.9\textwidth]{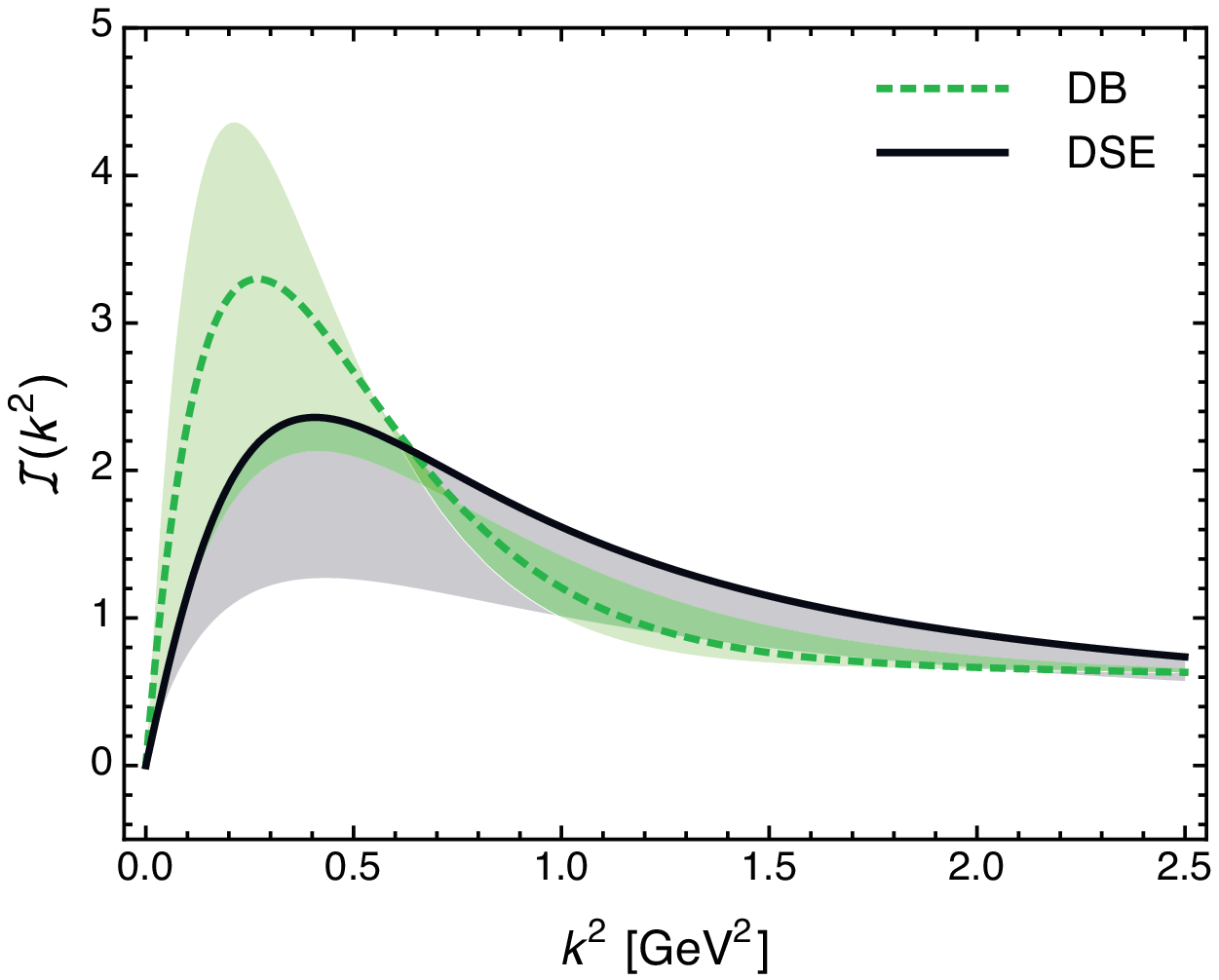}}
\end{minipage}
\begin{minipage}{0.47\textwidth}
\centerline{\includegraphics[clip,width=0.9\textwidth]{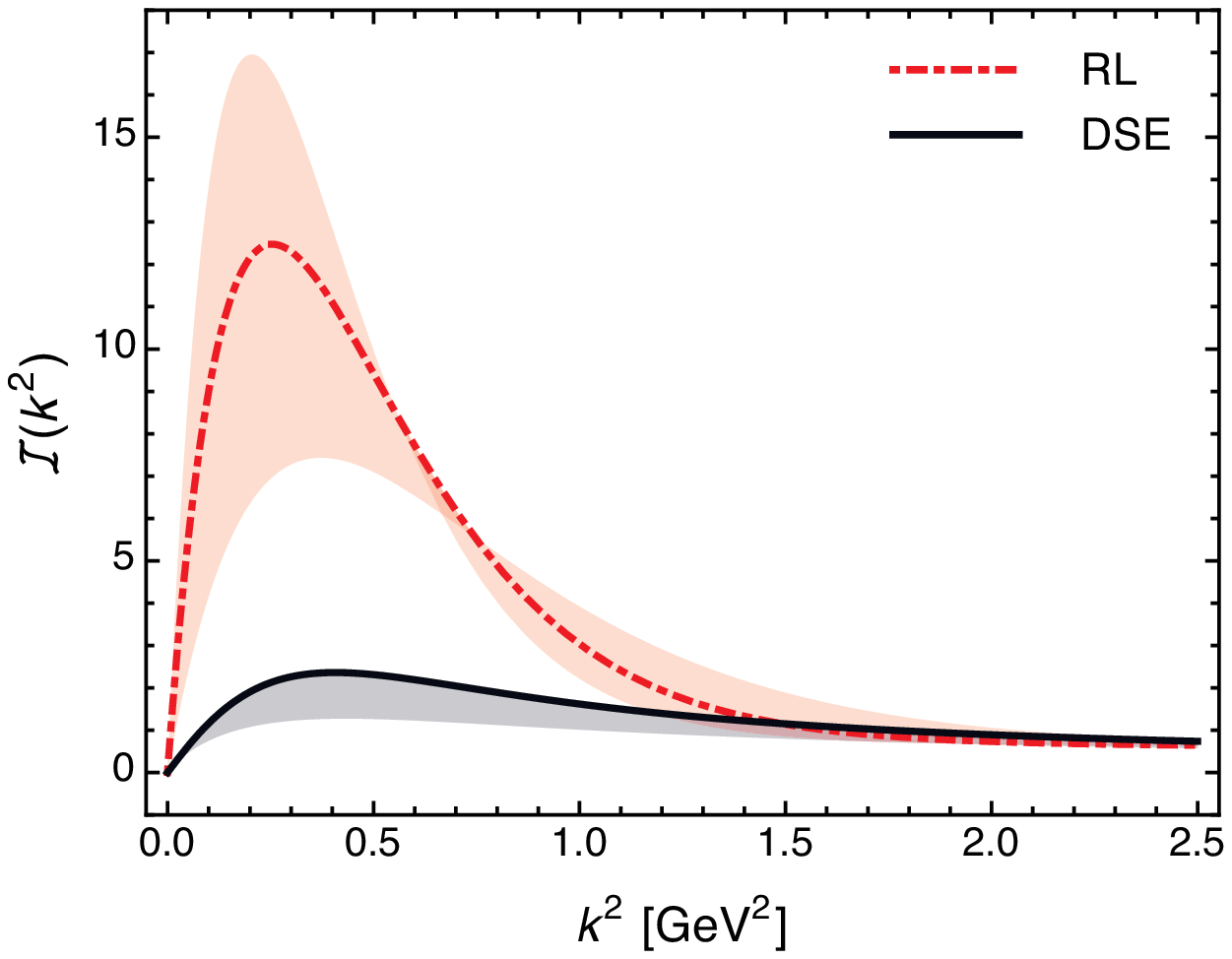}}
\end{minipage}
\end{minipage}}
\caption{Comparison between top-down results for the gauge-sector interaction and those obtained using the bottom-up approach based on hadron observables.
\textbf{Left panel} -- \emph{solid curve} within grey band, top-down result for the RGI interaction; and \emph{dashed curve} within \emph{pale-green band}, advanced bottom-up result obtained in the DB truncation \cite{Chang:2009zb, Chang:2010hb, Chang:2011ei}.
\textbf{Right panel} -- \emph{solid curve} within grey band, top-down result for the RGI interaction, as in the left panel; and \emph{dot-dashed curve} within \emph{pale-red band}, bottom-up result obtained in the RL truncation, which is  leading-order in the most widely-used DSE truncation scheme \cite{Binosi:2016rxzdFBS}.
In all cases, the bands denote the existing level of theoretical uncertainty in extraction of the interaction.
All curves are identical on the perturbative domain: $k^2>2.5\,$GeV$^2$.
(Figures provided by D.~Binosi, modelled after those in Ref.\,\cite{Binosi:2015xqkFBS}.)
\label{TopBottom}}
\end{figure}

Confidence in the insights drawn from continuum analyses of nonperturbative QCD has recently received a major boost owing to a unification of two common methods for determining the momentum-dependence of the interaction between quarks \cite{Binosi:2014aea}: namely, the top-down approach, which works toward an \textit{ab initio} computation of the interaction via direct analysis of the gauge-sector gap equations; and the bottom-up scheme, which aims to infer the interaction by fitting data within a well-defined truncation of those equations in the matter sector that are relevant to bound-state properties.  
The unification is illustrated in Fig.\,\ref{TopBottom}: the left panel presents a comparison between the top-down RGI interaction (solid-black curve) and the sophisticated DB-truncation bottom-up interaction (green band containing dashed curve).  Plainly, the interaction predicted by modern analyses of QCD's gauge sector is in near precise agreement with that required for a veracious description of measurable hadron properties using the most sophisticated matter-sector gap and Bethe-Salpeter kernels available today.  This is remarkable, given that there had previously been no serious attempt at communication between practitioners from the top-down and bottom-up hemispheres of continuum-QCD.  It bridges a gap that had lain between nonperturbative continuum-QCD and the \emph{ab initio} prediction of bound-state properties.

A comparison between the top-down prediction and that inferred using the simple DSE-RL kernel (red band containing dot-dashed curve in the right panel) is also important.  One observes that the DSE-RL result has the correct shape but is too large in the infrared.  This is readily explained \cite{Binosi:2014aea}; and it follows that whilst the RL truncation supplies a useful computational link between QCD's gauge sector and measurable hadron properties, the model interaction it delivers is \emph{not a pointwise-accurate} representation of ghost-gluon dynamics.  Notwithstanding this, it remains true that the judicious use of RL truncation can yield reliable predictions for a known range of hadron observables, with an error that may be estimated and whose origin is understood.

\section{Structure of Baryons}
This workshop focused on the electroproduction of nucleon resonances; and highlighted just how crucial it has become to address the three valence-quark bound-state problem in QCD with the same level of sophistication that is now available for mesons \cite{Chang:2011vu, Horn:2016rip}.  A principal modern goal must be to correlate the properties of meson and baryon ground- and excited-states within a single, \emph{symmetry-preserving} framework.  Here, symmetry-preserving means that the analysis respects Poincar\'e covariance and satisfies the relevant Ward-Green-Takahashi identities.  Constituent-quark models have hitherto been the most widely applied spectroscopic tools; and despite their imperfections, they are of continuing value because there is nothing yet that is providing a bigger picture.  Nevertheless, they possess no connection with quantum field theory; and they are not symmetry-preserving and hence cannot veraciously connect meson and baryon properties.

A comprehensive approach to QCD will provide a unified explanation of both mesons and baryons.  We have emphasised that DCSB is a keystone of the Standard Model, evident in the momentum-dependence of the dressed-quark mass function -- Fig.\,\ref{gluoncloud}; and it is just as important to baryons as it is to mesons.  Crucially, the DSEs furnish the only extant framework that can simultaneously and transparently connect meson and baryon observables with this basic feature of QCD, having provided, \emph{e.g}.\ a direct correlation of meson and baryon properties via a single interaction kernel, which preserves QCD's one-loop renormalisation group behaviour and can systematically be improved.  This is evident in  Refs.\,\cite{Eichmann:2008ae, Eichmann:2008ef, Eichmann:2011ej, Chang:2012cc, Segovia:2014aza, Roberts:2015dea, Segovia:2015hraS} and in many contributions to these proceedings, \emph{e.g}.\ Refs.\,\cite{Segovia:2016iaf, Eichmann:2016jqx, El-Bennich:2016qmb}.

In order to illustrate the insights that have been enabled by DSE analyses, consider the proton, which is a composite object whose properties and interactions are determined by its valence-quark content: $u$ + $u$ + $d$, \emph{i.e}.\ two up ($u$) quarks and one down ($d$) quark.  So far as is now known, bound-states seeded by two valence-quarks do not exist; and the only two-body composites are those associated with a valence-quark and -antiquark, \emph{i.e}.\ mesons.  These features are supposed to derive from colour confinement, whose complexities are discussed in the Introduction.

Such observations have led to a position from which the proton may be viewed as a Borromean bound-state \cite{Segovia:2015ufa, Segovia:2016iaf}, \emph{viz}.\ a system constituted from three bodies, no two of which can combine to produce an independent, asymptotic two-body bound-state.  In QCD the complete picture of the proton is more complicated, owing, in large part, to the loss of particle number conservation in quantum field theory and the concomitant frame- and scale-dependence of any Fock space expansion of the proton's wave function.  Notwithstanding that, the Borromean analogy provides an instructive perspective from which to consider both quantum mechanical models and continuum treatments of the nucleon bound-state problem in QCD.  It poses a crucial question:  \emph{Whence binding between the valence quarks in the proton, \mbox{\rm i.e}.\ what holds the proton together}?

In numerical simulations of lQCD that use static sources to represent the proton's valence-quarks, a ``Y-junction'' flux-tube picture of nucleon structure is produced \cite{Bissey:2006bz, Bissey:2009gw}.  This might be viewed as originating in the three-gluon vertex, which signals the non-Abelian character of QCD and is the source of asymptotic freedom.  Such results and notions would suggest a key role for the three-gluon vertex in nucleon structure \emph{if} they were equally valid in real-world QCD wherein light dynamical quarks are ubiquitous.  However, as explained in Sect.\,\ref{secConfinement}, they are not; and so a different explanation of binding within the nucleon must be found.

DCSB has numerous corollaries that are crucial in determining the observable features of the Standard Model; but one particularly important consequence is often overlooked.  Namely, any interaction capable of creating pseudo-Nambu-Goldstone modes as bound-states of a light dressed-quark and -antiquark, and reproducing the measured value of their leptonic decay constants, will necessarily also generate strong colour-antitriplet correlations between any two dressed quarks contained within a baryon.  This assertion is based upon evidence gathered in twenty years of studying two- and three-body bound-state problems in hadron physics.  No counter examples are known; and the existence of such diquark correlations is also supported by lQCD \cite{Alexandrou:2006cq, Babich:2007ahS}.

The properties of diquark correlations have been charted.  Most importantly, diquarks are confined.  Additionally, owing to properties of charge-conjugation, a diquark with spin-parity $J^P$ may be viewed as a partner to the analogous $J^{-P}$ meson \cite{Cahill:1987qr}.  It follows that scalar, isospin-zero and pseudovector, isospin-one diquark correlations are the strongest in ground-state $J^+$-baryons; and whilst no pole-mass exists, the following mass-scales, which express the strength and range of the correlation and are each bounded below by the partnered meson's mass, may be associated with these diquarks \cite{Cahill:1987qr, Maris:2002yu, Alexandrou:2006cq, Babich:2007ahS}:
$m_{[ud]_{0^+}} \approx 0.7-0.8\,$GeV, $m_{\{uu\}_{1^+}}  \approx 0.9-1.1\,$GeV,
with $m_{\{dd\}_{1^+}}=m_{\{ud\}_{1^+}} = m_{\{uu\}_{1^+}}$ in the isospin symmetric limit.
Realistic diquark correlations are also soft.  They possess an electromagnetic size that is bounded below by that of the analogous mesonic system, \emph{viz}.\ \cite{Maris:2004bp, Roberts:2011wyS}:
$r_{[ud]_{0^+}} \gtrsim r_\pi$, $r_{\{uu\}_{1^+}} \gtrsim r_\rho$,
with $r_{\{uu\}_{1^+}} > r_{[ud]_{0^+}}$.  As with mesons, these scales are set by that associated with DCSB.

The RGI interactions depicted in Fig.\,\ref{TopBottom} characterise a realistic class that generates strong attraction between two quarks and thereby produces tight diquark correlations in analyses of the three valence-quark scattering problem.  The existence of such correlations considerably simplifies analyses of baryon bound states because it reduces that task to solving a Poincar\'e covariant Faddeev equation \cite{Cahill:1988dx}, depicted in Fig.\,1, left panel, of Ref.\,\cite{Segovia:2016iaf}.  The three gluon vertex is not explicitly part of the bound-state kernel in this picture of the nucleon.  Instead, one capitalises on the fact that phase-space factors materially enhance two-body interactions over $n\geq 3$-body interactions and exploits the dominant role played by diquark correlations in the two-body subsystems.  Then, whilst an explicit three-body term might affect fine details of baryon structure, the dominant effect of non-Abelian multi-gluon vertices is expressed in the formation of diquark correlations.  Such a nucleon is then a compound system whose properties and interactions are primarily determined by its quark$+$diquark structure.

A nucleon (and any kindred baryon) with these features is a Borromean bound-state, the binding within which has two contributions.  One part is expressed in the formation of tight diquark correlations; but that is augmented by attraction generated by quark exchange (depicted in the shaded area of Fig.\,1, left panel, in Ref.\,\cite{Segovia:2016iaf}).  This exchange ensures that diquark correlations within the nucleon are fully dynamical: no quark holds a special place because each one participates in all diquarks to the fullest extent allowed by its quantum numbers. The continual rearrangement of the quarks guarantees, \emph{inter} \emph{alia}, that the nucleon's dressed-quark wave function complies with Pauli statistics.

One cannot overstate the importance of appreciating that these fully dynamical diquark correlations are vastly different from the static, pointlike ``diquarks'' which featured in early attempts \cite{Lichtenberg:1967zz} to understand the baryon spectrum and to explain the so-called missing resonance problem \cite{Ripani:2002ss, Burkert:2012ee, Kamano:2013iva}.  Modern diquarks are soft and enforce certain distinct interaction patterns for the singly- and doubly-represented valence-quarks within the proton, as reviewed in Refs.\,\cite{Segovia:2016iaf, Roberts:2015lja}.  On the other hand, the number of states in the spectrum of baryons obtained from the Faddeev equation \cite{Chen:2012qrS} is similar to that found in the three constituent-quark model, just as it is in today's lQCD spectrum calculations \cite{Edwards:2011jj}.

\section{Roper Resonance}
There was much discussion of the Roper resonance at this workshop.  That is unsurprising, given that the Roper has long resisted understanding.  Recently, however, JLab experiments \cite{Dugger:2009pn, Aznauryan:2009mx, Aznauryan:2011qj, Mokeev:2012vsa, Mokeev:2015lda, Burkert:2016dxc} have yielded precise nucleon-Roper ($N\to R$) transition form factors and thereby exposed the first zero seen in any hadron form factor or transition amplitude.  Additionally, Ref.\,\cite{Segovia:2015hraS} has provided the first continuum treatment of this problem using the power of relativistic quantum field theory.  A summary of that study is presented in these proceedings \cite{Segovia:2016iaf, El-Bennich:2016qmb}; and in this contribution, therefore, just a few points will be highlighted.

The analysis in Ref.\,\cite{Segovia:2015hraS} is distinguished by using dressed-quark propagators that express a running quark mass which connects textbook knowledge about QCD's ultraviolet behaviour with continuum- and lattice-QCD predictions for the mass function's infrared behaviour.  In addition, it capitalises on the existence of diquark correlations generated dynamically by the same mechanism that produces the dressed-quark mass function, \emph{viz}. DCSB.  These diquarks are soft, as predicted by QCD, and are active participants in all scattering processes.  They are also intrinsically active, \emph{i.e}. in a three-quark system, all diquarks participate in all correlations to the fullest extent allowed by their quantum numbers.  Finally, all these quantum field theoretical phenomena are combined via a Poincar\'e-covariant Faddeev equation and electromagnetic current in order to produce the nucleon and Roper masses and the $N\to R$ transition form factors, with no parameters varied, so that the results are true predictions upon which the framework must stand or fall.

A particular feature of Ref.\,\cite{Segovia:2015hraS} is that the computation yields only the contribution to the form factors from a rigorously defined dressed-quark core.  The mismatch between that result and data is then defined therein to be the effect of meson-baryon final-state interactions (MB-FSIs).  One may improve upon the estimate of MB-FSIs by recognising that the dressed-quark core component of the baryon Faddeev amplitudes should be renormalised by inclusion of meson-baryon ``Fock-space'' components, with a maximum strength of 20\% \cite{Cloet:2008fw, Bijker:2009up, Cloet:2014rja}.  Naturally, since wave functions in quantum field theory evolve with resolving scale, the magnitude of this effect is not fixed.  Instead ${\mathpzc I}_{MB}={\mathpzc I}_{MB}(Q^2)$, where $Q^2$ measures the resolving scale of any probe and ${\mathpzc I}_{MB}(Q^2) \to 0^+$ monotonically with increasing $Q^2$.  Now, form factors in QCD possess power-law behaviour, so it is reasonable to renormalise the dressed-quark core contributions via
\begin{equation}
\label{eqMBFSI}
F_{\rm core}(Q^2) \to [1- {\mathpzc I}_{MB}(Q^2)] F_{\rm core}(Q^2)\,,
\quad {\mathpzc I}_{MB}(Q^2) = [1-0.8^2]/[1+Q^2/\Lambda_{MB}^2]\,,
\end{equation}
with $\Lambda_{MB}=1\,$GeV marking the midpoint of the transition between the nonperturbative and perturbative domains of QCD as measured by the behaviour of the dressed-quark mass-function in Fig.\,\ref{gluoncloud}.

\begin{figure}[t]
\begin{minipage}[t]{\textwidth}
\begin{minipage}{0.47\textwidth}
\centerline{\includegraphics[clip,width=0.95\textwidth]{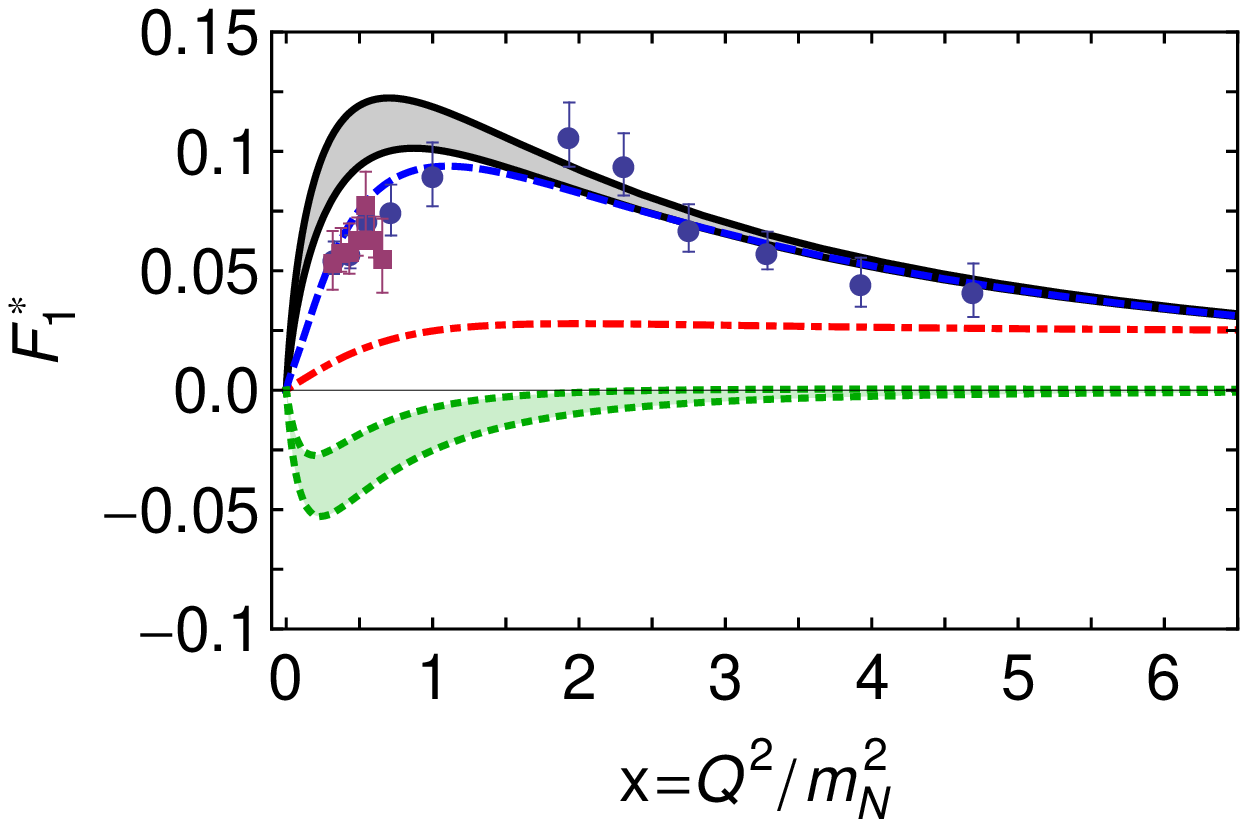}}
\end{minipage}
\begin{minipage}{0.47\textwidth}
\centerline{\includegraphics[clip,width=0.95\textwidth]{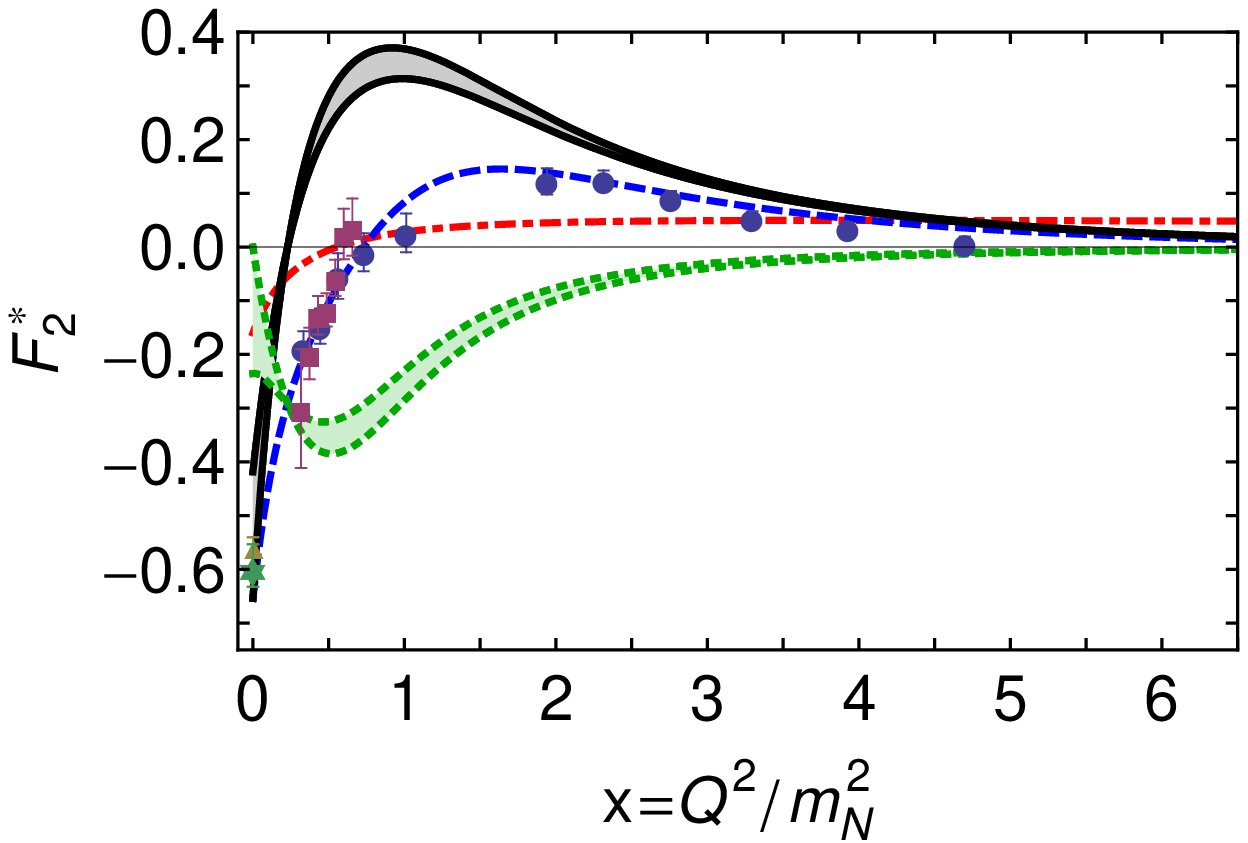}}
\end{minipage}
\end{minipage}
\begin{minipage}[t]{\textwidth}
\begin{minipage}{0.47\textwidth}
\centerline{\includegraphics[clip,width=0.95\textwidth]{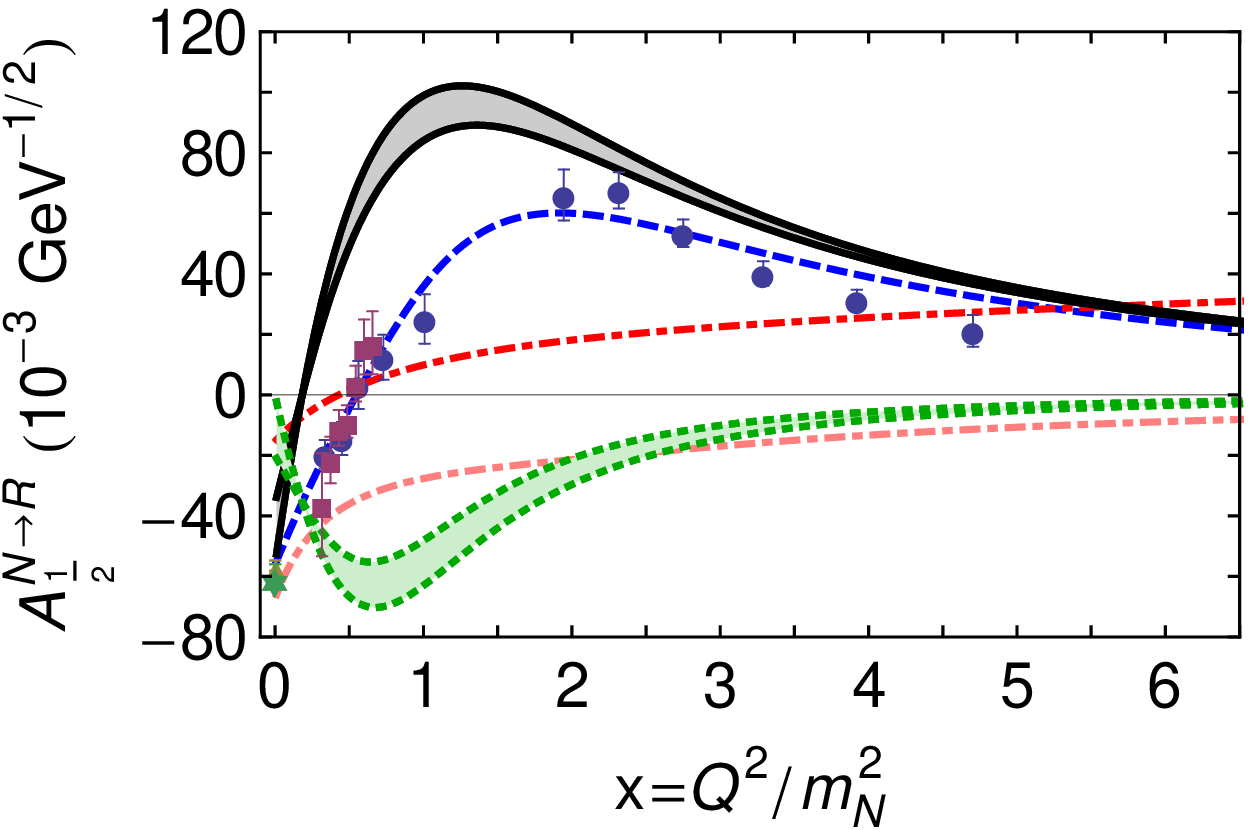}}
\end{minipage}
\begin{minipage}{0.47\textwidth}
\centerline{\includegraphics[clip,width=0.95\textwidth]{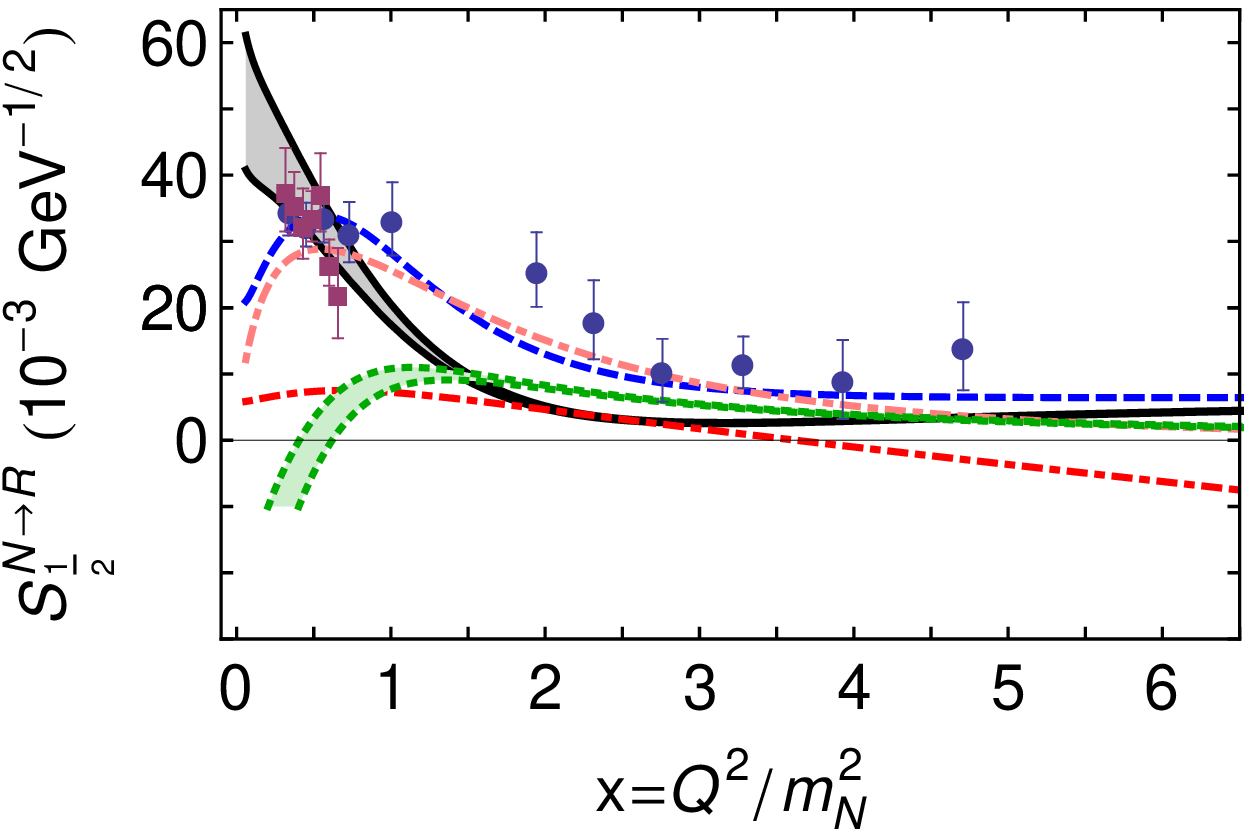}}
\end{minipage}
\end{minipage}
\caption{\label{MBcorrectedTFFs}
$N\to R$ transition form factors and helicity amplitudes obtained therefrom.  \emph{Legend}.  Grey band within black curves -- dressed-quark core contribution with up-to 20\% Faddeev amplitude renormalisation from MB-FSIs, implemented according to Eq.\,\eqref{eqMBFSI}.  The transition form factor curve with smallest magnitude at $x=1$ has the maximum renormalisation.
Green band within green dotted curves -- inferred MB-FSI contribution.  The band demarcates the range of uncertainty arising from $0\to 20$\% renormalisation of the dressed-quark core.
Blue dashed curve --  least-squares fit to the data on $x \in (0,5)$.
Red dot-dashed curve -- contact interaction result \cite{Wilson:2011aa}.
Pink dot-dashed curve, lower panels -- meson-cloud contribution to the helicity amplitude as determined by the excited baryon analysis center (EBAC).
Data: circles (blue) \cite{Aznauryan:2009mx};
triangle (gold) \cite{Dugger:2009pn};
squares (purple) \cite{Mokeev:2012vsa, Mokeev:2015lda};
and star (green) \cite{Agashe:2014kda}.}
\end{figure}

Following the procedure just described, one obtains the results depicted in Fig.\,\ref{MBcorrectedTFFs}, which highlight a number of important facts.
(\emph{i}) Incorporating a meson-baryon Fock-space component in the baryon Faddeev amplitudes does not materially affect the nature of the inferred meson-cloud contribution.
(\emph{ii}) Regarding $A_{1/2}$, the contribution of MB-FSIs inferred herein and that determined by EBAC are quantitatively in agreement on $x>1.5$.  However, our result disputes the EBAC suggestion that MB-FSIs are solely responsible for the $x=0$ value of the helicity amplitude: the quark-core contributes at least two-thirds of the result.
(\emph{iii}) Regarding $S_{1/2}$, our inference for the MB-FSIs raises serious questions about the EBAC result.  Indeed, the EBAC curve cannot realistically be connected with this helicity amplitude because there is necessarily a large quark-core contribution on $x<1$.  Moreover, the core and MB contributions are commensurate on $1<x<4$, and the core is dominant on $x>4$.
In our view, therefore, the green bands in these panels represent the best inference available today for the strength of MB-FSIs on the $N\to R$ transition form factors and helicity amplitudes.

One may summarise the main results of Ref.\,\cite{Segovia:2015hraS}, complemented by those presented herein, as follows. A range of properties of the dressed-quark core of the proton's first radial excitation were computed.  They provide an excellent understanding and description of data on the $N\to R$ transition and related quantities derived using dynamical coupled channels models.  The analysis is based on a sophisticated continuum framework for the three-quark bound-state problem; all elements employed possess an unambiguous link with analogous quantities in QCD; and no parameters were varied in order to achieve success.  Moreover, no material improvement in the results can be envisaged before either the novel spectral function methods introduced in Ref.\,\cite{Chang:2013pqS} are extended and applied to the entire complex of nucleon, $\Delta$-baryon and Roper-resonance properties that are unified by Refs.\,\cite{Segovia:2014aza, Roberts:2015dea, Segovia:2015hraS} or numerical simulations of lQCD become capable of reaching the same breadth of application and accuracy.  On the strength of these results and remarks one may confidently conclude that the observed Roper resonance is at heart the nucleon's first radial excitation and consists of a well-defined dressed-quark core augmented by a meson cloud that reduces its (Breit-Wigner) mass by approximately 20\%.  Concerning the transition form factors, a meson-cloud obscures the dressed-quark core from long-wavelength probes; but that core is revealed to probes with $Q^2 \gtrsim 3 m_N^2$.  This feature is typical of nucleon-resonance transitions; and hence measurements of resonance electroproduction on this domain can serve as an incisive probe of quark-gluon dynamics within the Standard Model, assisting greatly in mapping the evolution between QCD's nonperturbative and perturbative domains.

In connection with the last point and given the theme of this workshop, it is worth highlighting that, following completion of the JLab\,12 GeV upgrade, CLAS12 will be capable of determining the electrocouplings, $g_{vNN^\ast}(Q^2)$, of most prominent $N^\ast$ states at unprecedented photon virtualities: $Q^2\in [6,12]\,$GeV$^2$ \cite{E12-09-003, E12-06-108A}.  On this domain, Fig.\,\ref{MBcorrectedTFFs} suggests that these electrocouplings are primarily determined by the dressed-quark cores within baryons.  Consequently, the experimental programme employing CLAS12 will be unique in providing access to the dressed-quark cores of a diverse array of baryons.  It will therefore deliver empirical information that is necessary in order to address a wide range of critical issues, \emph{e.g}.: is there an environment sensitivity of DCSB and the dressed-quark mass function; and are quark-quark correlations an essential element in the structure of all baryons?  Existing feedback between experiment and theory indicates that there is no environment sensitivity for the nucleon, $\Delta$-baryon and Roper resonance \cite{Segovia:2014aza, Roberts:2015dea, Segovia:2015hraS}: DCSB in these systems is expressed in ways that can readily be predicted once its manifestation is understood in the pion, and this includes the generation of diquark correlations with the same character in each of these baryons.  Moreover, regarding the dressed-quark mass-function in Fig.\,\ref{gluoncloud}, the domain $Q^2\leq 12\,$GeV$^2$ translates into momenta $k \lesssim 1.2\,$GeV.  Therefore, combined CLAS and CLAS12 data are sensitive to the dressed-quark mass function throughout the domain upon which QCD dressing of quarks shifts from being essentially nonperturbative to perturbative in character.

\section{Conclusion}
It is worth reiterating a few points.
Owing to the conformal anomaly, both gluons and quarks acquire mass dynamically in QCD.  Those masses are momentum dependent, with large values at infrared momenta: $m(k^2\simeq 0) > \Lambda_{\rm QCD}$.
The appearance of these nonperturbative running masses is intimately connected with confinement and DCSB; and the relationship between those phenomena entails that in a Universe with light-quarks, confinement is a dynamical phenomenon.  Consequently, static-quark flux tubes are not the correct paradigm for confinement and it is practically meaningless to speak of linear potentials and Regge trajectories in connection with observable properties of light-quark hadrons.
In exploring the connection between QCD's gauge and matter sectors, top-down and bottom-up DSE analyses have converged on the form of the renormalisation-group-invariant interaction in QCD.  This outcome paves the way to parameter-free predictions of hadron properties.
Decades of studying the three valence-body problem in QCD have provided the evidence necessary to conclude that diquark correlations are a reality; but diquarks are complex objects, so their existence does not restrict the number of baryon states in any obvious way.  This effort has led to a sophisticated understanding of the nucleon, $\Delta$-baryon and Roper resonance: all may be viewed as Borromean bound-states, and the Roper is at heart the nucleon's first radial excitation.

The progress summarised herein highlights the capacity of DSEs in QCD to connect the quark-quark interaction, expressed, for instance, in the dressed-quark mass function, $M(p^2)$, with predictions for a wide range of hadron observables; and therefore serves as strong motivation for new experimental studies of, \emph{inter alia}, nucleon elastic and transition form factors, which exploit the full capacity of JLab\,12 in order to chart $M(p^2)$ and thereby explain the origin of more than 98\% of the visible mass in the Universe.
This must shed light on confinement, which is one of the most fundamental problems in modern physics and whose solution is unlikely to be found in a timely fashion through theoretical analysis alone.  A multipronged approach is required, involving constructive feedback between experiment and theory of the type illustrated herein.

\begin{acknowledgements}
Both the results described and the insights drawn herein are fruits from collaborations we have joined with many colleagues and friends throughout the world; and we are very grateful to them all.
We would also like to thank Ralf Gothe, Victor Mokeev and Elena Santopinto for enabling our participation in the ECT$^\ast$ Workshop: ``Nucleon Resonances: From Photoproduction to High Photon Virtualities'', 12-16 October 2015, which proved very rewarding.
This work was supported by the U.S.\ Department of Energy, Office of Science, Office of Nuclear Physics, under contract no.~DE-AC02-06CH11357; and the Alexander von Humboldt Foundation.
\end{acknowledgements}


\end{document}